\documentclass[aps,showpacs,twocolumn,nofootinbib]{revtex4-1}

\usepackage{graphicx}

\begin{document}

\title{The Higgs boson inclusive decay channels $H \to b\bar{b}$ and $H \to gg$ up to four-loop level}

\author{Sheng-Quan Wang}

\author{Xing-Gang Wu}
\email[email:]{wuxg@cqu.edu.cn}

\author{Xu-Chang Zheng}

\author{Jian-Ming Shen}

\author{Qiong-Lian Zhang}

\address{ Department of Physics, Chongqing University, Chongqing 401331, P.R. China}

\date{\today}

\begin{abstract}

The principle of maximum conformality (PMC) has been suggested to eliminate the renormalization scheme and renormalization scale uncertainties, which are unavoidable for the conventional scale setting and are usually important errors for theoretical estimations. In this paper, by applying PMC scale setting, we analyze two important inclusive Standard Model Higgs decay channels, $H\rightarrow b\bar{b}$ and $H\rightarrow gg$, up to four-loop and three-loop levels accordingly. After PMC scale setting, it is found that the conventional scale uncertainty for these two channels can be eliminated to a high degree. There is small residual initial scale dependence for the Higgs decay widths due to unknown higher-order $\{\beta_i\}$-terms. Up to four-loop level, we obtain $\Gamma(H\rightarrow b\bar{b}) = 2.389\pm0.073 \pm0.041$ MeV and up to three-loop level, we obtain $\Gamma(H\rightarrow gg) = 0.373\pm0.030$ MeV, where the first error is caused by varying $M_H=126\pm4$ GeV and the second error for $H\to b\bar{b}$ is caused by varying the $\overline{\rm MS}$-running mass $m_b(m_b)=4.18\pm0.03$ GeV. Taking $H\to b\bar{b}$ as an example, we present a comparison of three BLM-based scale setting approaches, e.g. the PMC-I approach based on the PMC-BLM correspondence, the $R_\delta$-scheme and the seBLM approach, all of which are designed to provide effective ways to identify non-conformal $\{\beta_i\}$-series at each perturbative order. At four-loop level, all those approaches lead to good pQCD convergence, they have almost the same pQCD series, and their predictions are almost independent on the initial renormalization scale. In this sense, those approaches are equivalent to each other. \\


\end{abstract}


\maketitle

\section{Introduction}

In 2012 a new boson has been discovered by CMS and ATLAS experiments at the Large Hadron Collider (LHC)~\cite{higgs1,higgs2}, whose properties are remarkably similar to the Standard Model (SM) Higgs~\cite{sign1,sign2,cms131,atlas131,atlascms131,higgsmass1}. For example, its mass is $\rm 125.5\pm0.2^{+0.5}_{-0.6}$ GeV by ATLAS collaboration~\cite{higgsmass1} or $\rm 125.7\pm0.3\pm0.3$ GeV by CMS collaboration~\cite{sign2}, where the first error stands for the statistic error and the second one stands for the systematic error.

If the SM Higgs has a mass around $126$ GeV, then its decay width shall be dominated by $H\rightarrow b\bar{b}$~\cite{exphbb,higgsrev}. Theoretically, many efforts have been made on studying the Higgs decays into a bottom pair~\cite{cor1,cor2,cor3,cor4,cor5,cor6,corbb,ew1,ew2,ew3,ew4,kat1,kat2,kat3}. As a reference, the pure QCD corrections at the two-loop or the three-loop level have been reported in Refs.~\cite{cor7,cor8,cor10}, and the explicit expressions up to $\mathcal{O}(\alpha^{5}_{s})$ have been given by Ref.~\cite{cor9}. In addition, the Higgs decay channel $H\rightarrow gg$ also plays a crucial role in studying the properties of Higgs boson. The coupling of Higgs to a pair of gluons, which is mediated at one loop by virtual massive quarks (essentially generated by the top quark alone) and becomes independent of the top-quark mass $m_{t}$ in the limit $M_{H} \ll 2m_{t}$. An important feature of $H\rightarrow gg$ is its affinity to the gluon-gluon fusion mechanism for Higgs production. The effective coupling $ggH$ eventually may provide a way to count the number of heavy quarks beyond SM~\cite{hevn}. Its next-to-leading order (NLO) QCD corrections are quite large and amount to about $70\%$ in comparison to the leading order contribution~\cite{hgg1,hgg2,hgg3,hgg4,hgg5,hgg6}. Later on, the QCD corrections for $H\rightarrow gg$ up to three-loop level have been calculated in the limit of an infinitely heavy top quark mass~\cite{hgg7,hgg8,hgg9,hgg10}. Those great improvements on loop calculations provide us chances for deriving more accurate estimation on Higgs properties.

The physical predictions of the theory, calculated up to all orders, are surely independent of the choice of renormalization scale and renormalization scheme due to the renormalization group invariance~\cite{peter1,bogo,peter2,callan,symanzik}. Whereas, as is well-known, there is renormalization scale and renormalization scheme ambiguities at any finite order. It is helpful to find an optimal scale setting so that there is no (or greatly suppressed) scale or scheme ambiguity at any fixed order and can achieve the most accurate estimation based on the known perturbative calculations.

As an estimation of the physical observable, one can first take an arbitrary initial renormalization scale $\mu_r=\mu_r^{\rm init}$ and apply some scale setting method to improve the pQCD estimation. For the conventional scale setting, once the renormalization scale is setting to be an initial value, it will always be fixed during the whole analysis. That is, for the processes involving Higgs, in the literature, one usually take $\mu_{r}\equiv\mu_r^{\rm init}=M_{H}$ as the central value, which eliminates the large logs in a form as $\ln(\mu_r/M_{H})$. Because there is no strong reasons for such a choice, as a compensation, one will vary the scale within a certain region, e.g. $\mu_r^{\rm init}\in [M_{H}/2, 2M_{H}]$, to ascertain the scale uncertainty. It is often argued that by setting and varying the scale in such a way, one can estimate contributions from higher order terms; i.e. changing in scale will affect how much of a result comes from Feynman diagrams without loops, and how much it comes from the leftover finite parts of loop diagrams. And because of its perturbative nature, it is a common belief that those scale uncertainties can be reduced after finishing a higher-and-higher order calculation. However, this {\it ad hoc} assignment of scale and its range usually constitutes an important systematic error in  theoretical and experimental analysis. More explicitly, the conventional scale setting can not answer the questions: why it is $M_{H}$ and not $M_{H}/2$ or others that provides the central estimation; when there are several typical energy scales for the process, then which one provides the central value ?

Several scale setting methods have been suggested, e.g. the renormalization group improved effective coupling method (FAC)~\cite{fac}, the principle of minimum sensitivity (PMS)~\cite{pms}, the Brodsky-Lepage-Mackenzie method (BLM)~\cite{blm} and its underlying principle of maximum conformality (PMC)~~\cite{pmc1,pmc11,pmc2,pmc3,pmc4,pmc5,pmc6,pmc7}. The FAC is to improve the perturbative series by requiring all higher-order terms vanish and the PMS is to force the fixed-order series to satisfy the renormalization group invariance at the renormalization point. The BLM improves the perturbative series by requiring the $n_f$-terms of at each perturbative order vanish. The PMC provides the principle underlying the BLM, and it suggests a principle to set the optimal renormalization scales up to all orders, they are equivalent to each other through the PMC - BLM correspondence principle~\cite{pmc2}. Those methods, being designed to eliminate the scale ambiguity, have quite different consequences and may or may not achieve their goals. A detailed introduction and comparison of these methods can be found in a recent review~\cite{pmc7}. In the present paper, we adopt PMC for analyzing the Higgs decays.

The main idea of PMC lies in that the PMC scales at each perturbative order are formed by absorbing all non-conformal terms that governs the running behavior of the coupling constant into the coupling constant. At each perturbative order, new types of $\{\beta_i\}$-terms will occur, so the PMC scale for each perturbative order is generally different. Even though, one can choose any value to be $\mu^{\rm init}_r$, the optimal PMC scales and the resulting finite-order PMC prediction are both to high accuracy independent of such arbitrariness. After PMC scale setting, the divergent renormalon series does not appear and the convergence of the pQCD series can be greatly improved. Because of these advantages, the PMC method can be widely applied to high energy physics processes, some examples of which can be found in Refs.~\cite{pmc2,pmc3,pmc4,pmc6,pmc11,jpsi,pom}.

Because the PMC provides the underlying principle for BLM, the previous features or properties derived by using BLM can also be understood by using PMC. Before applying PMC or BLM to high-energy processes, one needs to use the expression with full initial renormalization scale dependence. That is, those terms that have been eliminated by setting the renormalization scale to be equal to the factorization scale or by setting the initial renormalization to be the typical momentum transfer should be retrieved back. So, the previous scale dependence analysis or conclusions drawn under the BLM should be adopted with great care, since there is misuse of BLM in the literature. It is interesting to show whether the PMC can work well for the inclusive Higgs decays and whether the accuracy of the estimations can be improved. In present paper, we show the newly suggested PMC procedure, the so called $R_\delta$-scheme~\cite{pmc11,pmc12}, with much more detail. A comparison with some other suggestions to extend the BLM scale setting up to any perturbative orders shall also be presented.

The remaining parts of the paper are organized as follows. In Sec.II, we present the calculation technology for applying PMC to Higgs decay processes $H \rightarrow b\bar{b}$ and $H \rightarrow gg$ up to four-loop level. In Sec.III, we present the numerical results and discussions. The final section is reserved for a summary.

\section{Calculation technology}

In this section, we present an improved analysis for the Higgs decay channels $H \rightarrow b\bar{b}$ and $H \rightarrow gg$ by using the PMC $R_\delta$-scheme. For the purpose,
\begin{itemize}
\item We shall first rearrange the four-loop expressions~\cite{cor9,hgg9} that have been derived under the conventional scale setting in a more general form. That is, the $n_f$-terms in those expressions that are coming from the light-quark loops and are responsible for controlling the running behavior of the coupling constant shall be transformed into $\{\beta_i\}$-series. Those $\{\beta_i\}$-series via the renormalization group equation rightly control the running behavior of the strong coupling constant. Every process has its own $\{\beta_i\}$-series and its own optimal (PMC) scales. Thus, after absorbing all those $\{\beta_i\}$-series into the strong coupling constant via an order-by-order way, we can obtain the optimal running coupling constant for the specific process.

\item As stated in the Introduction, before applying the PMC scale setting, we need to obtain the expressions with full (initial) renormalization scale dependence. For the purpose, at present, we need to transform the four-loop results of $H\to b\bar{b}$ or the three-loop results of $H\to gg$ derived under the conventional assumption of $\mu^{\rm init}_r \equiv M_H$ to a more general form that explicitly contains the initial renormalization scale $\mu^{\rm init}_r$, which may or may not equal to $M_H$. This can be achieved by using the strong coupling constant's scale transformation equation up to four-loop level, i.e.
    \begin{widetext}
    \begin{eqnarray}
     a_{s}(Q^*) &=& a_{s}(Q)-\beta_{0} \ln\left(\frac{Q^{*2}}{Q^2}\right) a^{2}_{s}(Q) +\left[\beta^2_{0} \ln^2 \left(\frac{Q^{*2}}{Q^2}\right) -\beta_{1} \ln\left(\frac{Q^{*2}}{Q^2}\right) \right] a^{3}_{s}(Q) + \nonumber\\
     && \left[-\beta^3_{0} \ln^3 \left(\frac{Q^{*2}}{Q^2}\right) +\frac{5}{2} \beta_{0}\beta_{1} \ln^2\left(\frac{Q^{*2}}{Q^2}\right) -\beta_{2} \ln\left(\frac{Q^{*2}}{Q^2}\right)\right] a^{4}_{s}(Q) +{\cal O}(a^{5}_{s}) , \label{alphasrun}
     \end{eqnarray}
    \end{widetext}
    where $a_s=\alpha_s/4\pi$, $Q^*$ and $Q$ are two arbitrary renormalization scales.

\item We shall set the PMC scales in an order-by-order manner according to $R_\delta$-scheme. By doing the loop calculations, the $b$-quark mass is treated as massless, cf. the review on Higgs properties~\cite{higgsmb}, there is only an overall $m_b^2$-factor in the decay width. Since its value is irrelevant to the PMC procedures and should be kept separate during the PMC scale-setting, either the choice of $\overline{\rm MS}$-running mass or the pole mass is reasonable. In the formulae, we fix its value to be $m_{b}(M_H)$ within the $\overline{\rm MS}$-scheme. A detailed discussion on this point is in preparation, which shows by applying PMC properly, either the choice of pole mass or $\overline{\rm MS}$-running mass can get consistent estimation. Recently, a discussion on $H\to\gamma\gamma$ presents such an example~\cite{wang3}.
\end{itemize}

In the following, we sequently present the results before and after PMC scale setting for the two channels $H\to b\bar{b}$ and $H\to gg$.

\subsection{The general form for $H \rightarrow b\bar{b}$ under the conventional scale setting}

By taking the initial renormalization scale $\mu^{\rm init}_{r} = M_{H}$, the analytic decay width with explicit $n_f$ dependence for the channel $H\rightarrow b\bar{b}$ can be formally written as
\begin{widetext}
\begin{eqnarray}
\Gamma(H\rightarrow b\bar{b})&=& \frac{3G_{F}M_{H}m_{b}^{2}(M_{H})} {4\sqrt{2}\pi} \big[1+c_{1,0}\; a_{s}(M_{H})+ (c_{2,0}+c_{2,1}n_{f}) \; a_{s}^{2}(M_{H})+(c_{3,0}+c_{3,1}n_{f}+c_{3,2}n_{f}^{2}) \; a_{s}^{3}(M_{H}) \nonumber\\
&&\quad\quad\quad\quad\quad\quad\quad\quad\quad +(c_{4,0} +c_{4,1}n_{f}+c_{4,2}n_{f}^{2} +c_{4,3}n_{f}^{3})\; a_{s}^{4}(M_{H}) +{\cal O}(a_s^5) \big] \label{hbbnf} \\
&=& \frac{3G_{F}M_{H} m_{b}^{2}(M_{H})} {4\sqrt{2}\pi} \left[1+1.804\; \alpha_{s}(M_{H})+2.953\; \alpha_{s}^{2}(M_{H}) +1.347\; \alpha_{s}^{3}(M_{H}) - 8.475\; \alpha_{s}^{4}(M_{H}) +{\cal O}(\alpha_s^5)\right]
\end{eqnarray}
\end{widetext}
where $G_{F}$ is the Fermi constant and $m_{b}(M_{H})$ is the $\overline{\rm MS}$ running mass at the scale $M_H$. For convenience, in the third line, we present the values for the coefficients over the $\alpha_s$-expansion by setting $n_f=5$, which explicitly show the relative importance of the perturbative series. The coefficients $c_{i,j}$ up to four-loop levels are~\cite{cor9}
\begin{eqnarray}
c_{1,0}&=&22.667,\; c_{2,0}=575.04,\; c_{2,1}=-21.738, \nonumber\\
c_{3,0}&=&10504.9,\; c_{3,1}=-1649.4,\; c_{3,2}=16.574, \nonumber\\
c_{4,0}&=&10071,\; c_{4,1}=-56550,\; c_{4,2}=2479.4, \nonumber\\
c_{4,3}&=&-5.248.
\end{eqnarray}

The $R_\delta$-scheme~\cite{pmc11,pmc12} not only illuminates the $\{\beta_i\}$-pattern of the process but also exposes a special degeneracy of the coefficients at different perturbative orders. Such degeneracy is necessary, which, similar to the PMC - BLM correspondence principle~\cite{pmc2}, ensures the one-to-one correspondence between the $n_f$-series and the $\{\beta_i\}$-series at each perturbative order.

Applying $R_\delta$-scheme~\cite{pmc11,pmc12} to Eq.(\ref{hbbnf}), one can derive the general form of $H \rightarrow b\bar{b}$ for $\mu^{\rm init}_r \neq M_{H}$ with the help of Eq.(\ref{alphasrun}), which can be written as,
\begin{widetext}
\begin{eqnarray}
\Gamma(H\rightarrow b\bar{b})&=&\frac{3G_{F}M_{H}m_{b}^{2}(M_{H})} {4\sqrt{2}\pi} \bigg\{1+r_{1,0}(\mu^{\rm init}_{r})\; a_{s}(\mu^{\rm init}_{r}) + \left[ r_{2,0}(\mu^{\rm init}_{r})+\beta_{0}r_{2,1}(\mu^{\rm init}_{r})\right]\; a_{s}^{2}(\mu^{\rm init}_{r})\nonumber\\
&&+\left[ r_{3,0}(\mu^{\rm init}_{r})+\beta_{1}r_{2,1}(\mu^{\rm init}_{r})+2\beta_{0}r_{3,1}(\mu^{\rm init}_{r}) +\beta_{0}^{2}r_{3,2}(\mu^{\rm init}_{r})\right] \; a_{s}^{3}(\mu^{\rm init}_{r})+ \big[r_{4,0}(\mu^{\rm init}_{r})\nonumber\\
&&+\beta_{2}r_{2,1}(\mu^{\rm init}_{r})+2\beta_{1}r_{3,1}(\mu^{\rm init}_{r})+{5\over 2}\beta_{1}\beta_{0}r_{3,2}(\mu^{\rm init}_{r})+3\beta_{0} r_{4,1}(\mu^{\rm init}_{r})\nonumber\\
&&+3\beta_{0}^{2}r_{4,2}(\mu^{\rm init}_{r})+\beta_{0}^{3}r_{4,3}(\mu^{\rm init}_{r})\big] \; a_{s}^{4}(\mu^{\rm init}_{r}) + {\cal O}(a_{s}^{5}) \bigg\}. \label{hbbbt}
\end{eqnarray}
\end{widetext}
Here for later convenience, we have transformed the $n_f$-series into the required $\{\beta_i\}$-series. Explicit expressions for $\beta_0$, $\beta_1$ and $\beta_2$ in $n_f$-series can be found in Refs.\cite{beta1,beta2,beta3}. The $r_{i,0}$ with $i=(1,\cdots,4)$ are conformal coefficients, and the $r_{i,j}$ with $1\leq j<i\leq4$ are non-conformal coefficients that should be absorbed into the coupling constant. Those coefficients can be obtained by comparing Eq.(\ref{hbbnf}) with Eq.(\ref{hbbbt}), which are
\begin{eqnarray}
r_{1,0}(\mu^{\rm init}_{r})&=&c_{1,0};\\
r_{2,0}(\mu^{\rm init}_{r})&=&c_{2,0}+{33\over 2}c_{2,1}\\
r_{2,1}(\mu^{\rm init}_{r})&=&-{3\over 2}c_{2,1}-c_{1,0} \ln{M^{2}_{H}\over (\mu^{\rm init}_{r})^{2}}.\\
r_{3,0}(\mu^{\rm init}_{r})&=&{1\over 4} (-642c_{2,1}+4c_{3,0}+66c_{3,1}+1089c_{3,2}),\\
r_{3,1}(\mu^{\rm init}_{r})&=& {1\over 16} \Big(228c_{2,1}-12c_{3,1}-396c_{3,2}\nonumber\\
&&\left. -16c_{2,0}\ln{M^{2}_{H}\over (\mu^{\rm init}_{r})^{2}}-264c_{2,1}\ln{M^{2}_{H}\over (\mu^{\rm init}_{r})^{2}}\right), \\
r_{3,2}(\mu^{\rm init}_{r})&=&{9\over 4}c_{3,2} + c_{1,0}\ln^{2}{M^{2}_{H}\over (\mu^{\rm init}_{r})^{2}} + 3c_{2,1}\ln{M^{2}_{H}\over (\mu^{\rm init}_{r})^{2}}.\\
r_{4,0}(\mu^{\rm init}_{r})&=&{1\over 16}(11675c_{2,1}-2568c_{3,1}-84744c_{3,2}+ \nonumber\\
&&16c_{4,0}+264c_{4,1}+4356c_{4,2}+71874c_{4,3}),\\
r_{4,1}(\mu^{\rm init}_{r}) &=&{1\over16}\Big[-1916c_{2,1} +152c_{3,1}+8226c_{3,2}-8c_{4,1} \nonumber\\
&&-264c_{4,2}-6534c_{4,3}+2568c_{2,1}\ln{M^{2}_{H}\over (\mu^{\rm init}_{r})^{2}} \nonumber\\
&&-16c_{3,0}\ln{M^{2}_{H}\over (\mu^{\rm init}_{r})^{2}}-264c_{3,1}\ln{M^{2}_{H}\over (\mu^{\rm init}_{r})^{2}}\nonumber\\
&&-4356c_{3,2}\ln{M^{2}_{H}\over (\mu^{\rm init}_{r})^{2}}\Big], \\
r_{4,2}(\mu^{\rm init}_{r})&=&{1\over 16}\left({325\over 3}c_{2,1}-570c_{3,2}+12c_{4,2}+594c_{4,3}\right.\nonumber\\
&&\left. +16c_{2,0}\ln^{2}{M^{2}_{H}\over (\mu^{\rm init}_{r})^{2}}+264c_{2,1}\ln^{2}{M^{2}_{H}\over (\mu^{\rm init}_{r})^{2}}\right. \nonumber\\
&&\left. -456c_{2,1}\ln{M^{2}_{H}\over (\mu^{\rm init}_{r})^{2}}+24c_{3,1}\ln{M^{2}_{H}\over (\mu^{\rm init}_{r})^{2}} \right. \nonumber\\
&&\left. +792c_{3,2}\ln{M^{2}_{H}\over (\mu^{\rm init}_{r})^{2}}\right), \\
r_{4,3}(\mu^{\rm init}_{r})&=&{1\over 6}\left(-{81\over 4}c_{4,3}-6c_{1,0}\ln^{3}{M^{2}_{H}\over (\mu^{\rm init}_{r})^{2}}- \right. \nonumber\\
&&\left. 27c_{2,1}\ln^{2}{M^{2}_{H}\over (\mu^{\rm init}_{r})^{2}}-{81\over 2}c_{3,2}\ln{M^{2}_{H}\over (\mu^{\rm init}_{r})^{2}}\right).
\end{eqnarray}
It is noted that, as expected, the conformal coefficients $r_{i,0}$ are independent of $\mu_r^{\rm init}$, and we can omit the variable $\mu^{\rm init}_r$ in $r_{i,0}$. \\

\subsection{Results for $H\to b\bar{b}$ after PMC scale setting}

Following the standard procedure of the $R_\delta$-scheme, by absorbing all non-conformal $\{\beta_i\}$-series that control the running behavior of the coupling constant into the coupling constant, we can reduce Eq.(\ref{hbbbt}) to the following conformal series,
\begin{widetext}
\begin{eqnarray}
\Gamma(H\rightarrow b\bar{b})&=& \frac{3G_{F}M_{H} m_{b}^{2}(M_{H})} {4\sqrt{2}\pi} \left[1+r_{1,0} a_{s}(Q_{1})+r_{2,0} a_{s}^{2}(Q_{2})+r_{3,0} a_{s}^{3}(Q_{3}) +r_{4,0} a_{s}^{4}(Q_{4})\right] \nonumber\\
&=& \frac{3G_{F}M_{H} m_{b}^{2}(M_{H})} {4\sqrt{2}\pi} \left[1+1.804\; \alpha_{s}(Q_{1})+1.370\; \alpha_{s}^{2}(Q_{2}) -4.389\; \alpha_{s}^{3}(Q_{3}) - 4.430\; \alpha_{s}^{4}(Q_{4})\right] ,
\end{eqnarray}
\end{widetext}
where in the second line, we present the values for the conformal coefficients over the $\alpha_s$ expansion, which show the relative importance of the perturbative series. Here $Q_{i}$ with $i=(1,\cdots,4)$ are PMC scales, which can be obtained through the following formulas
\begin{widetext}
\begin{eqnarray}
Q_{1} &=& \mu^{\rm init}_{r}\exp\left\{ \frac{1}{2} \frac{-r_{2,1}(\mu_{r}^{\rm init})+{r_{3,2}(\mu_{r}^{\rm init})\over 2}{\partial{\beta}\over \partial{a_{s}}}-{r_{4,3}(\mu_{r}^{\rm init})\over3!} \left[\beta{\partial{^{2}\beta}\over \partial{a_{s}^{2}}}+\left({\partial{\beta}\over \partial{a_{s}}}\right)^{2}\right] }{r_{1,0}(\mu_{r}^{\rm init})-{r_{2,1}(\mu_{r}^{\rm init})
\over 2}\left({\partial{\beta}\over \partial{a_{s}}}\right)+{r_{3,2}(\mu_{r}^{\rm init})\over 4}\left({\partial{\beta}\over \partial{a_{s}}}\right)^{2}+{1\over 3!} \left[\beta{\partial{^{2}\beta}\over \partial{a_{s}^{2}}}-{1\over 2}\left({\partial{\beta}\over \partial{a_{s}}}\right)^{2}\right] {r_{2,1}^{2}(\mu_{r}^{\rm init})\over r_{1,0}(\mu_{r}^{\rm init})}}\right\}, \label{pmcq1} \\
Q_{2} &=& \mu^{\rm init}_{r} \exp\left\{\frac{1}{2} \frac{-r_{3,1}(\mu_{r}^{\rm init})+{r_{4,2}(\mu_{r}^{\rm init})\over 2}\left[{\partial{\beta}\over \partial{a_{s}}}+{\beta \over a_{s}}\right]} { r_{2,0}(\mu_{r}^{\rm init})-{r_{3,1}(\mu_{r}^{\rm init})\over 2}
\left[{\partial{\beta}\over \partial {a_{s}}}+{\beta \over a_{s}}\right]}\right\}, \label{pmcq2} \\
Q_{3} &=& \mu^{\rm init}_{r} \exp\left\{ \frac{1}{2} \frac{-{r_{4,1}(\mu_{r}^{\rm init})}}{r_{3,0}(\mu_{r}^{\rm init})}\right\}, \label{pmcq3}
\end{eqnarray}
\end{widetext}
where $\beta=-a_{s}^{2}\sum\limits_{i=0}^{\infty}\beta_{i}a_{s}^{i}$ being the conventional QCD renormalization group $\beta$-function. We note that the last
scale leaves some ambiguity in PMC scale setting, since there is no $\{\beta_i\}$-terms that can determine its optimal value, we set $Q_{4}=\mu^{\rm init}_r$. Because the PMC scales ($Q_1$, $Q_2$ and $Q_3$) themselves are in perturbative series, the residual scale dependence due to unknown higher-order $\{\beta_i\}$-terms shall be highly suppressed.

\subsection{The decay width of the $H\rightarrow gg$}

By taking the initial renormalization scale $\mu^{\rm init}_{r} = M_{H}$, the analytic expression for the decay width of $H\rightarrow gg$ can be written as
\begin{widetext}
\begin{eqnarray}
\Gamma(H\rightarrow gg)&=& {4G_{F}M^{3}_{H} \over 9\sqrt{2}\pi} \big[c_{1,0} \; a^{2}_{s}(M_{H})+(c_{2,0}+c_{2,1}n_{f}) \; a_{s}^{3}(M_{H})+(c_{3,0} +c_{3,1}n_{f}+c_{3,2}n_{f}^{2}) \; a_{s}^{4}(M_{H})\nonumber\\
&&\quad\quad\quad\quad\quad\quad\quad\quad\quad\quad +(c_{4,0} +c_{4,1}n_{f}+c_{4,2}n_{f}^{2} +c_{4,3}n_{f}^{3}) \; a_{s}^{5}(M_{H}) +{\cal O}(a_s^6) \big] \label{hggnf} \\
&=& {4G_{F}M^{3}_{H} \over 9\sqrt{2}\pi}\times10^{-3} \left[6.333\;\alpha_s^2(M_{H}) + 36.114\;\alpha_s^3(M_{H}) -98.267\;\alpha_s^4(M_{H})+ 80.443\;\alpha_s^5(M_{H}) + {\cal O}(\alpha_s^6)\right] .
\end{eqnarray}
\end{widetext}
For convenience, in the third line, we present the values for the coefficients over the $\alpha_s$-expansion by setting $n_f=5$, which explicitly show the relative importance of the perturbative series. The coefficients $c_{i,j}$ are~\cite{hgg9}
\begin{eqnarray}
c_{1,0}&=&1; c_{2,0}=95.0, c_{2,1}=-4.667;\nonumber\\
c_{3,0}&=&5898.8, c_{3,1}=-761.81, c_{3,2}=14.428;\nonumber\\
c_{4,0}&=&287583, c_{4,1}=-68580, c_{4,2}=3393.1, c_{4,3}=-34.42.\nonumber\\
\end{eqnarray}

From Eq.(\ref{hggnf}), the general form for the Higgs decay process $H \rightarrow gg$ with $\mu^{\rm init}_{r} \neq M_{H}$ can be written as the following form,
\begin{widetext}
\begin{eqnarray}
\Gamma(H\rightarrow gg)&=&\frac{4G_{F}M^{3}_{H}}{9\sqrt{2}\pi} \big\{ r_{1,0}(\mu_{r}^{\rm init}) a^{2}_{s}(\mu_{r}^{\rm init}) +[r_{2,0}(\mu_{r}^{\rm init})+2\beta_{0}r_{2,1}(\mu_{r}^{\rm init})] a_{s}^{3}(\mu_{r}^{\rm init}) + [r_{3,0}(\mu_{r}^{\rm init})+2\beta_{1}r_{2,1}(\mu_{r}^{\rm init})\nonumber\\
&&+3\beta_{0}r_{3,1}(\mu_{r}^{\rm init})+3\beta_{0}^{2}r_{3,2}(\mu_{r}^{\rm init})] a_{s}^{4}(\mu_{r}^{\rm init}) +[r_{4,0}(\mu_{r}^{\rm init}) +2\beta_{2}r_{2,1}(\mu_{r}^{\rm init})+3\beta_{1}r_{3,1}(\mu_{r}^{\rm init}) \nonumber\\
&&+7\beta_{1}\beta_{0}r_{3,2}(\mu_{r}^{\rm init}) +4\beta_{0}r_{4,1}(\mu_{r}^{\rm init})+6\beta_{0}^{2}r_{4,2}(\mu_{r}^{\rm init})+4\beta_{0}^{3}r_{4,3}(\mu_{r}^{\rm init})] a_{s}^{5}(\mu_{r}^{\rm init}) +{\cal O}(a^6_s) \big\}. \label{hggbt}
\end{eqnarray}
\end{widetext}
Following the same procedures of $R_\delta$-scheme, the conformal or non-conformal coefficients $r_{i,j}(\mu^{\rm init}_{r})$ can be written as
\begin{eqnarray}
r_{1,0}(\mu^{\rm init}_{r})&=&c_{1,0}.\\
r_{2,0}(\mu^{\rm init}_{r})&=&c_{2,0}+{33\over 2}c_{2,1}, \\
r_{2,1}(\mu^{\rm init}_{r})&=&-{3\over 4}c_{2,1}-c_{1,0}\ln{M^{2}_{H}\over (\mu^{\rm init}_{r})^{2}}. \\
r_{3,0}(\mu^{\rm init}_{r})&=&{1\over 4}(-642c_{2,1}+4c_{3,0}+66c_{3,1}+1089c_{3,2}), \\
r_{3,1}(\mu^{\rm init}_{r})&=&{1\over 24}\Big(228c_{2,1}-12c_{3,1}-396c_{3,2} \nonumber\\
&&-396c_{2,1}\ln{M^{2}_{H}\over (\mu^{\rm init}_{r})^{2}}-24c_{2,0}\ln{M^{2}_{H}\over (\mu^{\rm init}_{r})^{2}}\Big), \\
r_{3,2}(\mu^{\rm init}_{r})&=&\frac{3}{4}c_{3,2} + c_{1,0}\ln^{2}{M^{2}_{H} \over (\mu^{\rm init}_{r})^{2}} + \frac{3}{2} c_{2,1}\ln{M^{2}_{H}\over (\mu^{\rm init}_{r})^{2}}. \\
r_{4,0}(\mu^{\rm init}_{r})&=&{1\over 8}(-12721c_{2,1}-1284c_{3,1} -42372c_{3,2}+ \nonumber\\
&&8c_{4,0}+132c_{4,1}+2178c_{4,2}+35937c_{4,3}),  \\
r_{4,1}(\mu^{\rm init}_{r}) &=& {1\over32} \Big(-1416c_{2,1}+ 228c_{3,1}+12018c_{3,2}-12c_{4,1}\nonumber\\
&&-396c_{4,2}-9801c_{4,3}+5136c_{2,1}\ln{M^{2}_{H}\over (\mu^{\rm init}_{r})^{2}} \nonumber \\
&&-32c_{3,0}\ln{M^{2}_{H}\over (\mu^{\rm init}_{r})^{2}} -528c_{3,1}\ln{M^{2}_{H}\over (\mu^{\rm init}_{r})^{2}}\nonumber\\
&&-8712c_{3,2}\ln{M^{2}_{H}\over (\mu^{\rm init}_{r})^{2}}\Big), \\
r_{4,2}(\mu^{\rm init}_{r})&=&{1\over 48} \Big(325c_{2,1}-798c_{3,2} +18c_{4,2}+891c_{4,3}\nonumber\\
&&+48c_{2,0}\ln^{2}{M^{2}_{H}\over (\mu^{\rm init}_{r})^{2}} +792c_{2,1}\ln^{2}{M^{2}_{H}\over (\mu^{\rm init}_{r})^{2}}\nonumber\\
&&-912c_{2,1}\ln{M^{2}_{H}\over (\mu^{\rm init}_{r})^{2}} +48c_{3,1}\ln{M^{2}_{H}\over (\mu^{\rm init}_{r})^{2}}\nonumber\\
&&+1584c_{3,2}\ln{M^{2}_{H}\over (\mu^{\rm init}_{r})^{2}}\Big),\\
r_{4,3}(\mu^{\rm init}_{r})&=&-{1\over 32}\left(27c_{4,3}+ 32c_{1,0}\ln^{3}{M^{2}_{H}\over (\mu^{\rm init}_{r})^{2}}\right. \nonumber\\
&&\left. +72c_{2,1}\ln^{2}{M^{2}_{H}\over (\mu^{\rm init}_{r})^{2}}+72c_{3,2}\ln{M^{2}_{H}\over (\mu^{\rm init}_{r})^{2}} \right).
\end{eqnarray}
It is noted that, as required, the conformal coefficients $r_{i,0}$ are independent of $\mu_r^{\rm init}$ and we can omit the argument $(\mu^{\rm init}_r)$ in $r_{i,0}$.

By absorbing all non-conformal $\{\beta_i\}$-series that control the running behavior of the coupling constant into the coupling constant, we can reduce Eq.(\ref{hggbt}) into the following conformal series,
\begin{widetext}
\begin{eqnarray}
\Gamma(H\rightarrow gg)&=&{4G_{F}M^{3}_{H} \over 9\sqrt{2}\pi} \left[r_{1,0} a^{2}_{s}(Q_{1}) +r_{2,0} a_{s}^{3}(Q_{2}) + r_{3,0} a_{s}^{4}(Q_{3}) +r_{4,0} a_{s}^{5}(Q_{4}) + {\cal O}(a^6_s) \right] \nonumber \\
&=& {4G_{F}M^{3}_{H} \over 9\sqrt{2}\pi}\times10^{-3} \left[6.333\;\alpha_s^2(Q_1) + 9.068\;\alpha_s^3(Q_2) -79.962\;\alpha_s^4(Q_3)- 68.804\;\alpha_s^5(Q_4) + {\cal O}(\alpha_s^6)\right] ,
\end{eqnarray}
\end{widetext}
where in the second line, we present the values for the conformal coefficients over the $\alpha_s$ expansion. Here $r_{i,0}(\mu_{r}^{\rm init})$ are conformal coefficients. The PMC scales $Q_i$ with $i=(1,2,3)$ can be obtained from the same way as that of Eqs.(\ref{pmcq1},\ref{pmcq2},\ref{pmcq3}), $Q_4$ is also the undetermined scale due to the unknown higher order $\{\beta_i\}$ terms and we also fix it to be $\mu^{\rm init}_r$.

\section{Numerical results and discussions} \label{Numer}

To do numerical calculation, we take $G_{F}=1.16638\times10^{-5}\rm GeV^{-2}$, the Higgs mass $M_{H}=126$ GeV and the top quark mass $m_{t}=173.5$ GeV. We adopt the four-loop $\alpha_{s}$ running with its asymptotic scale determined by the fixed point $\alpha_s(M_{Z})=0.1184$~\cite{pdg}: $\Lambda_{\rm QCD}^{n_{f}=3}=0.339 $ GeV, $\Lambda_{\rm QCD}^{n_{f}=4}=0.296 $ GeV and $\Lambda_{\rm QCD}^{n_{f}=5}=0.213 $ GeV. The $\overline{\rm MS}$-running quark mass ${m}_{b}(m_b)=4.18\pm0.03$ GeV~\cite{pdg}, and by using the quark mass anomalous dimension expressions listed in Ref.\cite{mb1,mb2}, we obtain $m_b(M_H)=2.78 \pm0.02 $ GeV.

\subsection{Basic results for $H\rightarrow b\bar{b}$}

\begin{table*}[ht]
\begin{center}
\begin{tabular}{|c||c|c|c|c|c|c||c|c|c|c|c|c|}
\hline
& \multicolumn{6}{c||}{Conventional scale setting} & \multicolumn{6}{c|}{PMC scale setting} \\
\hline\hline
~~~ ~~~ & ~LO~ & ~NLO~ & ~N$^{2}$LO~ & ~N$^{3}$LO~ &~N$^{4}$LO~ &~Total~ & ~LO~& ~NLO~& ~N$^{2}$LO~ & ~N$^{3}$LO~ & ~N$^{4}$LO~ & ~Total~ \\
\hline
$\Gamma_{i}$ (KeV) & 1924.28 & 391.74  & 72.38 &3.73& -2.65 & 2389.48 & 1924.28 &436.23 &48.12 & -18.12 & -1.38 & 2389.13\\
\hline
$\Gamma_{i}/\Gamma_{\rm tot}$ ~&80.53\%&16.39\%&3.03\%&0.16\%&-0.11\%& &80.54\%&18.26\%&2.01\%&-0.76\%&-0.06\%&\\
\hline
\end{tabular}
\caption{Decay width for $H\rightarrow b\bar{b}$ up to four-loop level. For conventional scale setting, we set the renormalization scale $\mu_r\equiv M_H$. For PMC scale setting, we set the initial renormalization scale $\mu^{\rm init}_r=M_H$. Here $\Gamma_i$ stands for the decay width at each perturbative order with $i={\rm LO}$, ${\rm NLO}$ and etc., $\Gamma_{\rm tot}$ stands for the total decay width. $M_H=126$ GeV. } \label{tablep}
\end{center}
\end{table*}

The decay widths of $H\rightarrow b\bar{b}$ before and after PMC scale setting are presented in Table \ref{tablep}, where $\Gamma_i$ stands for the decay width at each perturbative order with $i={\rm LO}$, ${\rm NLO}$ and etc., and $\Gamma_{\rm tot}$ stands for the total decay width. We set the renormalization scale $\mu_r \equiv\mu^{\rm init}_r = M_H$ for the conventional scale setting, and we take the initial renormalization scale $\mu^{\rm init}_r=M_H$ to initialize the PMC scale setting.

Table \ref{tablep} shows that, either before or after PMC scale setting, about $80\%$ contribution comes from the LO order terms, which is exact and free from the strong interactions. The total decay width for $H\rightarrow b\bar{b}$ remains almost unchanged, $\Gamma(H\to b\bar{b})\simeq 2.39$ MeV. This shows the choice of $\mu_r\equiv M_H$ is a lucky guess for the conventional scale setting. Because of the elimination of the renormalon terms, one could expect a better pQCD convergence after PMC scale setting. This is clearly shown in Table \ref{tablep}, e.g. the four-loop terms only give $\sim 0.1\%$ contributions to the total decay width.

\begin{figure}[h]
\includegraphics[width=0.50\textwidth]{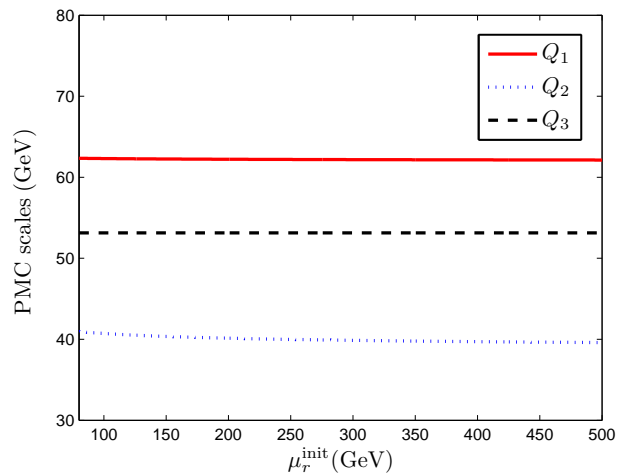}
\caption{The LO, NLO and NNLO PMC scales $Q_{1}$, $Q_{2}$ and $Q_3$ versus the initial renormalization scale $\mu^{\rm init}_r$ for the $H\rightarrow b\bar{b}$ process, which are shown by solid, dotted and dashed lines, respectively. $M_H=126$ GeV. }
\label{Plot:pmcscale}
\end{figure}

For the $H\rightarrow b\bar{b}$ process at $\mathcal{O}(\alpha_{s}^{4})$, we need to introduce four PMC scales, i.e. the LO PMC scale $Q_1$, the NLO PMC scale $Q_2$, the N$^2$LO PMC scale $Q_3$ and the N$^3$LO PMC scale $Q_4$. As stated in the last section, since there is no $\beta$-terms that can determine its optimal value, we set $Q_{4} \equiv \mu^{\rm init}_r$ \footnote{This corresponds to the second type of residual scale dependence after PMC scale setting~\cite{pom}, which, as is the present case, can also be highly suppressed when the pQCD convergence is under well control. }. Using the formulas (\ref{pmcq1},\ref{pmcq2},\ref{pmcq3}), we show how the PMC scales depend on the initial renormalization scale, which are presented in FIG.(\ref{Plot:pmcscale}). FIG.(\ref{Plot:pmcscale}) shows that the PMC scales $Q_{1,2,3}$ are highly independent on the choice of initial renormalization scale. This indicates that the PMC scale setting do provide a principle for setting the optimal (solitary) renormalization scale of high energy processes. For example, setting $\mu_{r}^{\rm init}=M_{H}$, we obtain
\begin{equation}
Q_{1}=62.3 \;{\rm GeV},\; Q_{2}=40.5 \; {\rm GeV}, \; Q_{3}=53.1 \; {\rm GeV} .
\end{equation}
These PMC scales are smaller than $M_{H}$ to a certain degree due to the exponential suppressions from the absorbtion of higher order $\{\beta_i\}$-terms. These PMC scales are different, which shows that they are controlled by different $\{\beta_i\}$-series at each perturbative order.

\begin{figure}[t]
\includegraphics[width=0.50\textwidth]{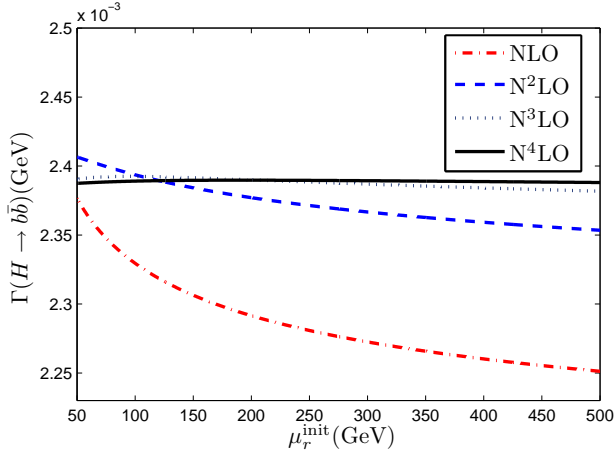}
\caption{Total decay width $\Gamma(H\to b\bar{b})$ up to four-loop level under conventional scale setting versus the scale $\mu_r\equiv\mu^{\rm init}_r$. The dash-dot, dashed, dotted and solid lines are for NLO, N$^2$LO, N$^3$LO and N$^4$LO estimations, respectively. }\label{Plot:hbb}
\end{figure}

\begin{figure}[t]
\includegraphics[width=0.50\textwidth]{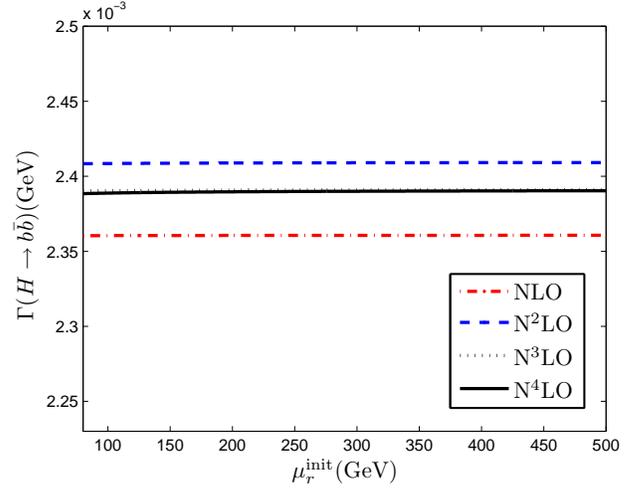}
\caption{Total decay width $\Gamma(H\to b\bar{b})$ up to four-loop level after PMC scale setting versus the initial renormalization scale $\mu^{\rm init}_r$. The dash-dot, dashed, dotted and solid lines are for NLO, N$^2$LO, N$^3$LO and N$^4$LO estimations, respectively. }\label{Plot:hbbpmc}
\end{figure}

As a further comparison, we show the total decay width $\Gamma(H\to b\bar{b})$ versus the initial renormalization scale $\mu^{\rm init}_{r}$ before and after PMC scale setting in FIGs.(\ref{Plot:hbb},\ref{Plot:hbbpmc}). In these two figures, the dash-dot, dashed, dotted and solid lines are for NLO, N$^2$LO, N$^3$LO and N$^4$LO estimations, respectively. From these two figures, we observe that
\begin{itemize}
\item FIG.(\ref{Plot:hbb}) interprets the idea of the conventional scale setting. Conventionally, the renormalization scale is taken as the typical momentum transfer of the process or a value which minimizes the contributions of the loop diagrams. For the present process, $\mu_r\equiv\mu^{\rm init}_r = M_H$. The total decay width $\Gamma(H\to b\bar{b})$ shows a relatively strong dependence on the value of $\mu^{\rm init}_r$ at the NLO level, e.g.
    \begin{eqnarray}
    \mu^{\rm init}_r &=& M_H/2 \to \Gamma(H\to b\bar{b})\simeq 2.36 \; {\rm MeV}, \nonumber\\
    \mu^{\rm init}_r &=& M_H \to \Gamma(H\to b\bar{b})\simeq 2.32 \; {\rm MeV}, \nonumber\\
    \mu^{\rm init}_r &=& 2M_H \to \Gamma(H\to b\bar{b})\simeq 2.28 \; {\rm MeV}, \nonumber\\
    \mu^{\rm init}_r &=& 4M_H \to \Gamma(H\to b\bar{b})\simeq 2.25 \; {\rm MeV}.
    \end{eqnarray}
    This shows that the NLO scale error is $\pm2\%$ for $\mu^{\rm init}_r\in[M_H/2,2M_H]$ and $\left(^{+2\%}_{-3\%}\right)$ for $\mu^{\rm init}_r\in[M_H/2,4M_H]$. As one includes higher-and-higher orders, the guessed scale will lead to a better estimate. For example, at the NNLO level, we have
    \begin{eqnarray}
    \mu^{\rm init}_r &=& M_H/2 \to \Gamma(H\to b\bar{b})\simeq 2.40 \; {\rm MeV}, \nonumber\\
    \mu^{\rm init}_r &=& M_H \to \Gamma(H\to b\bar{b})\simeq 2.39 \; {\rm MeV}, \nonumber\\
    \mu^{\rm init}_r &=& 2M_H \to \Gamma(H\to b\bar{b})\simeq 2.37 \; {\rm MeV}, \nonumber\\
    \mu^{\rm init}_r &=& 4M_H \to \Gamma(H\to b\bar{b})\simeq 2.35 \; {\rm MeV}.
    \end{eqnarray}
    This shows that the NNLO scale error reduces to $\left(^{+0.5\%}_{-0.8\%}\right)$ for $\mu^{\rm init}_r\in[M_H/2,2M_H]$ and $\left(^{+0.5\%}_{-1.7\%}\right)$ for $\mu^{\rm init}_r\in[M_H/2,4M_H]$. When considering up to three-loop level or four-loop level, the decay width becomes almost invariant within the present considered region of $[M_H/2,4M_H]$. This agrees with the conventional wisdom that by finishing a higher enough calculation, one can get desirable convergent and scale-invariant estimations.

    We would like to stress that even if a proper choice of $\mu^{\rm init}_r$ may lead to a value close to the experimental data by using conventional scale setting, this may not be the correct answer for a fixed-order estimation. Especially, if a process does not converge enough, one has to finish a more and more complex loop calculations so as to achieve the precision goal.

\item FIG.(\ref{Plot:hbbpmc}) shows that after PMC scale setting, the total decay width of $H\rightarrow b\bar{b}$ are almost flat versus the choice of renormalization scale even at the NLO level. This is due to the fact that the PMC scales $Q_{1}$, $Q_{2}$ and $Q_{3}$ themselves are highly independent on the choice of $\mu^{\rm init}_r$, as shown by FIG.(\ref{Plot:pmcscale}). The residual scale dependence due to unknown higher-order $\{\beta_i\}$-series has been highly and exponentially suppressed.

    It is noted that there is slight difference for the decay widths at different perturbative orders, e.g.
    \begin{eqnarray}
    \Gamma(H\to b\bar{b})|_{\rm NLO} &\simeq& 2.36 \;{\rm MeV} , \nonumber\\
    \Gamma(H\to b\bar{b})|_{\rm N^2LO} &\simeq& 2.41 \;{\rm MeV}, \nonumber\\
    \Gamma(H\to b\bar{b})|_{\rm N^3LO} &\simeq& 2.39 \;{\rm MeV}, \nonumber\\
    \Gamma(H\to b\bar{b})|_{\rm N^4LO} &\simeq& 2.39 \;{\rm MeV}. \nonumber
    \end{eqnarray}
    Such difference shows that even though by absorbing the non-conformal terms into the coupling constant, one can greatly improve the pQCD convergence and simultaneously eliminate the scale dependence at lower perturbative orders, one may still need to know higher-order conformal contributions if one wants to achieve even higher precision. For examples, the N$^2$LO improves NLO estimation by about $2\%$ and the N$^3$LO improves N$^2$LO estimation by about $1\%$. More over, the unknown higher-order non-conformal contributions can be roughly estimated by varying the final undetermined PMC scale as $Q_4$ via the conventional way, e.g. $[Q_4/2,2Q_4]$.

\end{itemize}

\subsection{Basic results for $H\rightarrow gg$}

We can estimate the properties of $H\rightarrow gg$ in a similar way as that of $H\rightarrow b\bar{b}$.

\begin{table*}[ht]
\begin{center}
\begin{tabular}{|c||c|c|c|c|c||c|c|c|c|c|}
\hline
& \multicolumn{5}{c||}{Conventional scale setting} & \multicolumn{5}{c|}{PMC scale setting} \\
\hline
~~~ ~~~ & ~~LO~~ & ~~NLO~~ & ~~N$^{2}$LO~~ & ~~N$^{3}$LO~~ & ~~Total~~ & ~~LO~~& ~~NLO~~ & ~~N$^{2}$LO~~ & ~~N$^{3}$LO~~ & ~~Total~~ \\
\hline
$\Gamma_{i}$ (KeV) & 188.27 & 121.18  & 37.21 &3.26 & 349.92 & 332.36 &117.84 &-74.45 & -2.94 & 372.81\\
\hline
$\Gamma_{i}/\Gamma_{tot}$ ~&53.80\%&34.63\%&10.63\%&0.93\%& &89.15\%&31.61\%&-20.00\%&-0.79\%&\\
\hline
\end{tabular}
\caption{Decay width for the process $H\rightarrow gg$ up to three-loop level. For conventional scale setting, we set the renormalization scale $\mu_r\equiv M_H$. For PMC scale setting, we set the initial renormalization scale $\mu^{\rm init}_r=M_H$. Here $\Gamma_i$ stands for the decay width at each perturbative order with $i={\rm LO}$, ${\rm NLO}$ and etc., $\Gamma_{\rm tot}$ stands for the total decay width. $M_H=126$ GeV. } \label{tablepc}
\end{center}
\end{table*}

We put the decay width for $H\rightarrow gg$ before and after PMC scale setting in Table \ref{tablepc}, where $\Gamma_i$ stands for the decay width at each perturbative order with $i={\rm LO}$, ${\rm NLO}$ and etc., and $\Gamma_{\rm tot}$ stands for the total decay width. The total decay width $\Gamma_{\rm tot}$ is $\sim 350$ KeV under the conventional scale setting, which improves to be $\sim 373$ KeV after PMC scale setting. Such a small increment in some sense means the choice of $\mu_{r}^{\rm init}=M_{H}$ is a viable choice for the conventional scale setting up to three-loop level. Under the conventional scale setting, we have
\begin{eqnarray}
&& \frac{\Gamma_{\rm LO}}{\Gamma_{\rm tot}} : \frac{\Gamma_{\rm NLO}}{\Gamma_{\rm tot}} : \frac{\Gamma_{\rm N^2LO}}{\Gamma_{\rm tot}} : \frac{\Gamma_{\rm N^3LO}}{\Gamma_{\rm tot}} \nonumber\\
&\approx& 54\%: 35\% : 11\% : 0.9\%
\end{eqnarray}
and after PMC scale setting, we have
\begin{eqnarray}
&& \frac{\Gamma_{\rm LO}}{\Gamma_{\rm tot}} : \frac{\Gamma_{\rm NLO}}{\Gamma_{\rm tot}} : \frac{\Gamma_{\rm N^2LO}}{\Gamma_{\rm tot}} : \frac{\Gamma_{\rm N^3LO}}{\Gamma_{\rm tot}} \nonumber\\
&\approx& 89\%: 32\% : (-20\%) : 0.8\% .
\end{eqnarray}
This shows that for the decay of $H\to gg$, only after a three-loop correction, one can obtain a desirable convergent estimation. Note the pQCD convergence after PMC is weaker than the case for the conventional scale setting for $H\to gg$, since the ${\rm N^2LO}$ part becomes $(-20\%)$. This could mean that we need to know an more accurate $\{\beta_i\}$ series so as to determine the ${\rm N^2LO}$ PMC scale.

\begin{figure}[ht]
\includegraphics[width=0.50\textwidth]{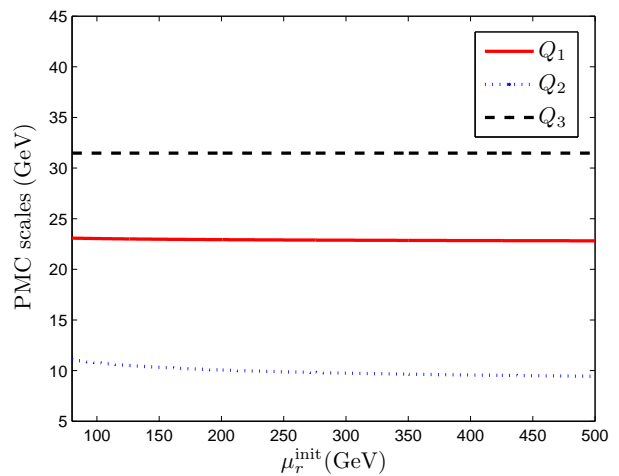}
\caption{The LO, NLO and NNLO PMC scales $Q_{1}$, $Q_{2}$ and $Q_3$ versus the initial renormalization scale $\mu^{\rm init}_r$ for the $H\rightarrow gg$ process, which are shown by solid, dotted and dashed lines, respectively. $M_H=126$ GeV. } \label{Plot:hggpmcscale}
\end{figure}

\begin{figure}[ht]
\includegraphics[width=0.50\textwidth]{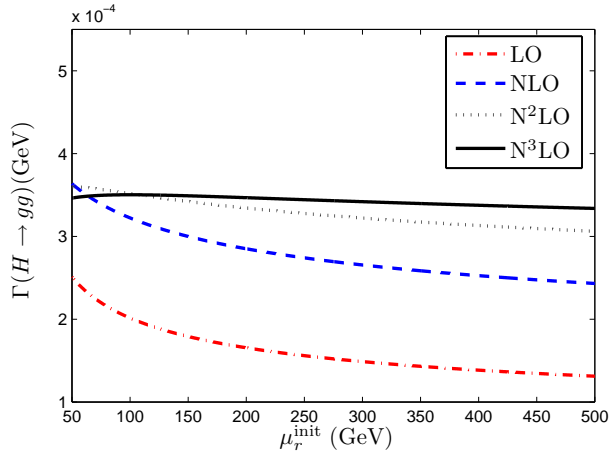}
\caption{Total decay width $\Gamma(H\to gg)$ up to three-loop level under conventional scale setting versus the renormalization scale $\mu_r\equiv\mu^{\rm init}_r$. The dash-dot, dashed, dotted and solid lines are for LO, NLO, N$^2$LO and N$^3$LO estimations, respectively. } \label{Plot:hgg}
\end{figure}

\begin{figure}[ht]
\includegraphics[width=0.50\textwidth]{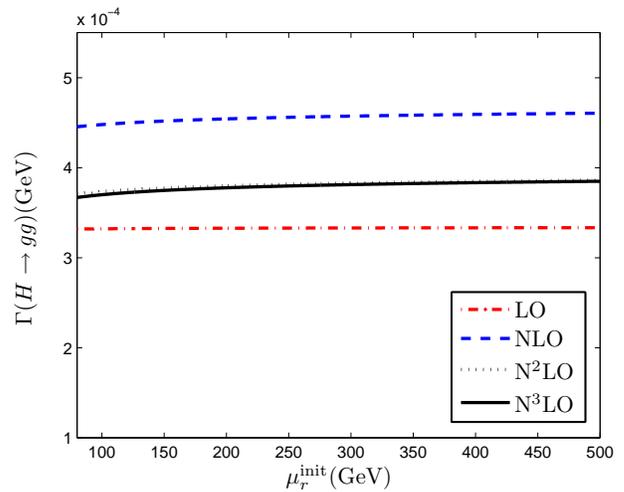}
\caption{Total decay width $\Gamma(H\to gg)$ up to three-loop level after PMC scale setting versus the initial renormalization scale $\mu^{\rm init}_r$. The dash-dot, dashed, dotted and solid lines are for LO, NLO, N$^2$LO and N$^3$LO estimations, respectively. } \label{Plot:hggpmc}
\end{figure}

For the $H\rightarrow gg$ process at $\mathcal{O}(\alpha_{s}^{5})$, we need to introduce four PMC scales, i.e. the LO PMC scale $Q_1$, the NLO PMC scale $Q_2$, the N$^2$LO PMC scale $Q_3$ and the N$^3$LO PMC scale $Q_4$. As stated in the last section, since there is no $\beta$-terms that can determine its optimal value, we set $Q_{4} \equiv \mu^{\rm init}_r$. FIG.(\ref{Plot:hggpmcscale}) shows that the PMC scales $Q_{1,2,3}$ are highly independent of the choice of initial renormalization scale. This indicates that the PMC scale setting do provide a principle for setting the optimal (solitary) renormalization scale of high energy processes. By setting $\mu_{r}^{\rm init}=M_{H}$ GeV, we obtain
\begin{equation}
Q_{1}=23.0\; {\rm GeV},\; Q_{2}=10.5\;{\rm GeV},\; Q_{3}=31.5 \; {\rm GeV}.
\end{equation}
These PMC scales are also smaller than $M_{H}$ to a certain degree due to exponential suppressions from the absorbtion of higher order $\{\beta_i\}$-terms. We show the total decay width $\Gamma(H\to gg)$ versus the initial renormalization scale $\mu^{\rm init}_{r}$ before and after PMC scale setting in FIGs.(\ref{Plot:hgg},\ref{Plot:hggpmc}). In these two figures, the dash-dot, dashed, dotted and solid lines are for LO, NLO, N$^2$LO and N$^3$LO estimations, respectively.

These results show that for the decay channel of $H\rightarrow gg$, we can obtain similar conclusions as those of $H\rightarrow b\bar{b}$. More explicitly,
\begin{itemize}
\item FIG.(\ref{Plot:hgg}) indicates that, under the conventional scale setting, the total decay width $\Gamma(H\to gg)$ shows a strong dependence on $\mu^{\rm init}_r$ at the LO level, e.g.
    \begin{eqnarray}
    \mu^{\rm init}_r &=& M_H/2 \to \Gamma(H\to gg)\simeq 232 \; {\rm KeV}, \nonumber\\
    \mu^{\rm init}_r &=& M_H \to \Gamma(H\to gg) \simeq 188 \; {\rm KeV}, \nonumber\\
    \mu^{\rm init}_r &=& 2M_H \to \Gamma(H\to b\bar{b})\simeq 156 \; {\rm KeV}, \nonumber\\
    \mu^{\rm init}_r &=& 4M_H \to \Gamma(H\to b\bar{b})\simeq 131 \; {\rm KeV}.
    \end{eqnarray}
    This shows that the LO scale error is $40\%$ for $\mu^{\rm init}_r\in[M_H/2,2M_H]$ and $54\%$ for $\mu^{\rm init}_r\in[M_H/2,4M_H]$. Moreover, the scale error for $\mu^{\rm init}_r\in[M_H/2,4M_H]$ shall change down to $35\%$, $15\%$ and $4\%$ for NLO, N$^2$LO and N$^3$LO estimations, respectively. This shows that as one includes higher-and-higher orders, the guessed scale will lead to a better estimation. For the $H\to gg$ decay, only after a three-loop correction, one can obtain a desirable small about several percent scale error.

\item FIG.(\ref{Plot:hggpmc}) shows that, after PMC scale setting, the total decay widths of $H\rightarrow gg$ up to the mentioned perturbative orders are almost flat versus the choice of initial renormalization scale $\mu^{\rm init}_r$. The residual scale dependence due to unknown higher-order $\{\beta_i\}$-series has been highly suppressed. Similar to the case of $H\to b\bar{b}$, this is due to the fact that the PMC scales $Q_{1}$, $Q_{2}$ and $Q_{3}$ themselves are highly independent on the choice of $\mu^{\rm init}_r$, which are shown clearly by FIG.(\ref{Plot:hggpmcscale}).

\end{itemize}

\subsection{A comparison of the approaches underlying BLM scale setting}

The BLM scale setting is designed to improve the pQCD predictions by absorbing the $n_f$-terms via a proper way into the coupling constant~\cite{blm}. Since its invention by Brodsky-Lepage-Mackenzie in 1983, the BLM has been widely accepted in the literature for dealing with high energy processes, such as the $e^+e^-\to $ hadrons, the deep inelastic scattering, the heavy meson or baryon productions or decays, the exclusive processes such as the pion-photon transition form factors, the QCD lattice perturbative theory, and etc.. Encouraged by its great successes, several approaches have been tried to extend BLM to any perturbative orders or put it in a more solid background, such as the PMC-I approach (first approach to achieve the goal of PMC via the PMC-BLM correspondence principle)~\cite{pmc1,pmc2}, the $R_\delta$-scheme (second approach to achieve the goal of PMC)~\cite{pmc11,pmc12} and the seBLM approach~\cite{seblm,seblm2}.

It is noted that the role of the running coupling in any gauge theory is to absorb the physics of the $\beta$ function, which governs its running behavior via the renormalization group equations. Any approach that properly identifies the $\{\beta_i\}$-series for a physical observable will lead to equivalently the same result. Practically, one usually calculate the $n_f$-terms by considering the vacuum polarization contributions. However, different ways of identifying $n_f$-series to $\{\beta_i\}$-series may lead to: I) different effective $\{\beta_i\}$-series at each known perturbative order; II) different residual $\{\beta_i\}$-dependence because of unknown perturbative orders; III) different conformal terms leaving at each perturbative order; IV) and different pQCD convergence. If one can do the perturbative corrections to a higher enough perturbative order, different effective schemes may result in consistent physical predictions. The equivalence of the PMC-I approach and $R_\delta$-scheme have already been shown in Refs.\cite{pmc11,pmc12,pmc7}. In the following, we take the Higgs decay channel $H\to b\bar{b}$ as an explicit example to show that the seBLM approach is also consistent with the PMC approaches. One subtle point of such a comparison (an also any applications of those approaches) lies in that, we should first transform the estimations with full renormalization scale dependence with the help of the transformation formulae (\ref{alphasrun}).

In PMC-I approach, by introducing an PMC-BLM correspondence principle in which the $\{\beta_i\}$-series for a physical observable has the same parton of the running coupling itself~\cite{pmc2}, the number of the effective independent $\{\beta_i\}$-terms exactly corresponds to the number of $n_f$-terms at each perturbative order. In $R_\delta$-scheme, by introducing the ``degeneracy" properties of the $\{\beta_i\}$-series observed by a generalization of the conventional dimensional regularization scheme to any dimensional-like ones~\cite{pmc11}, one can also obtain a one-to-one correspondence between the $\{\beta_i\}$-series and the $n_f$-series. The calculation technologies for those two self-consistent approaches can be found in the corresponding references, the interesting readers may turn to those references or very recent review~\cite{pmc7} for detail. In Sec.II, we have presented our analysis under the $R_\delta$-scheme.

\begin{table*}[ht]
\begin{center}
\begin{tabular}{|c||c|c|c|c|c|c|c|}
\hline
  & ~~$\Gamma_{\rm tot}$~~ & ~~$\Gamma_{\rm LO}/\Gamma_{\rm tot}$~~ & ~~$\Gamma_{\rm NLO}/\Gamma_{\rm tot}$~~ & ~~$\Gamma_{\rm N^{2}LO}/\Gamma_{\rm tot}$~~ & ~~$\Gamma_{\rm N^3LO}/\Gamma_{\rm tot}$~~ & ~~$\Gamma_{\rm N^4LO}/\Gamma_{\rm tot}$~~ \\
\hline
~~conventional scale setting~~ & ~~ 2.389 MeV ~~ & ~~ 80.53\% ~~ & ~~ 16.39\%  ~~ & ~~ 3.03\% ~~ & ~~ 0.16\% ~~ & ~~ $-0.11$\%~~  \\
\hline
~~seBLM~~ & ~~2.389 MeV~~ & ~~ 80.56\% ~~ & ~~ 18.25\% ~~ & ~~ 1.99\% ~~ & ~~ -0.72\% ~~ & ~~ -0.08\%~~ \\
\hline
~~PMC-I~~& ~~2.388 MeV~~ & ~~80.58\%~~ &~~18.26\%~~&~~2.03\%~~&~~-0.76\%~~&~~-0.10\%~~\\
\hline
~~$R_\delta$-scheme~~ & ~~2.389 MeV~~ &~~ 80.54\% ~~&~~ 18.26\% ~~&~~ 2.01\% ~~&~~ -0.76\% ~~&~~ -0.06\%~~ \\
\hline
~~BKM~\cite{BKM01}~~ & ~~2.75 MeV ~~&~~ 74.5\% ~~&~~ 17.7\%  ~~&~~ 5.3\% ~~&~~ 1.8\% ~~&~~ 0.7\%\\
\hline
~~FAPT with $l=2$~\cite{mfapt}~~ & ~~2.38 MeV~~ &~~ 79.5\% ~~&~~ 16.2\%  ~~&~~ 4.3\% ~~&~~ - ~~&~~ - ~~ \\
\hline
~~FAPT with $l=3$~\cite{mfapt}~~ & ~~2.44 MeV~~ &~~ 78.5\% ~~&~~ 16.1\%  ~~&~~ 4.2\% ~~&~~ 1.2\% ~~&~~ - ~~ \\
\hline
\end{tabular}
\caption{A comparison of several approaches for calculating the perturbative coefficients of $H\rightarrow b\bar{b}$, where the predictions of the PMC-I scheme, the $R_\delta$-scheme and the seBLM scheme, together with the ones derived under conventional scale setting, are presented. Here $\Gamma_i$ stands for the decay width at each perturbative order with $i={\rm LO}$, ${\rm NLO}$ and etc., $\Gamma_{\rm tot}$ stands for the total decay width. The initial renormalization scale is taken as $M_{H}=126$ GeV. To be useful reference, the results of Refs.\cite{mfapt,BKM01} for the FAPT scheme and the BKM scheme are also presented. }
\label{tableoth}
\end{center}
\end{table*}

While the seBLM scheme provides quite a different way from those two PMC approaches, in which a general $\{\beta_i\}$-series at each perturbative order have been introduced, and in order to get an one-to-one correspondence between the $n_f$-series with the $\{\beta_i\}$-series, some extra approximations (or equivalently some extra degrees of freedom) have to be introduced~\cite{seblm}. More explicitly, the seBLM scheme transforms the standard power series $a^{n}_{s}(\mu_{r}^{\rm init})$ to the series of the products $\prod_{i=1}^{n}a_{s}(Q_{i})$. After applying the seBLM scheme to Eq.(\ref{hbbnf}), the decay width of the process $H\to b\bar{b}$ can be expressed as follows:
\begin{widetext}
\begin{displaymath}
\Gamma(H\rightarrow b\bar{b}) = {3G_{F}M_{H} \over 4\sqrt{2}\pi} m_{b}^{2}(M_{H}) \left[1+{r'_{1,0}(\mu_{r}^{\rm init})\over \beta_{0}}A_{1}\bigg(1+{A_{2}\over \beta_{0 }} \bigg(r'_{2,0}(\mu_{r}^{\rm init}) +{A_{3}\over \beta_{0 }} \bigg(r'_{3,0}(\mu_{r}^{\rm init})+{A_{4}\over \beta_{0 }}(r'_{4,0}(\mu_{r}^{\rm init})+...)\bigg)\bigg)\bigg)\right],
\end{displaymath}
\end{widetext}
where $r'_{i}$ stands for the conformal coefficients of seBLM leaving at each perturbative order, $A_{i}=\beta_{0}a_{s}(Q_{i})$ stand for the redefined coupling constant. Because we have no higher-order $\{\beta_{i}\}$-terms to determine the scale for $A_{4}$, we set $Q_{4}=Q_{3}$ as suggested by seBLM. Three effective seBLM scales are
\begin{eqnarray}\label{seblmscale}
\ln\left({(\mu_{r}^{\rm init})^2 \over Q^2_{1}}\right)&=& \Delta_{1,0}+\Delta_{1,1}A_{1} +\Delta_{1,2}A_{1}^{2}, \label{seblmscale1}\\
\ln\left({Q^2_{1}\over Q^2_{2}}\right) &=& \Delta_{2,0}+\Delta_{2,1} A_{2},\label{seblmscale2}\\
\ln\left({Q^2_{2}\over Q^2_{3}}\right) &=& \Delta_{3,0}.\label{seblmscale3}
\end{eqnarray}
where the explicit expressions of the coefficients $\Delta_{i,j}$ can be found in Ref.\cite{seblm} \footnote{Because for $H\to b\bar{b}$, we have no extra constraints or degrees of freedom to set an one-to-one correspondence between the $n_f$-series and the $\{\beta_i\}$-series, as an estimation, we directly adopt the same pattern of $\{\beta_i\}$-series in each perturbative order that has been derived for Adler D-function for the present process~\cite{seblm}. }. Then, we obtain
\begin{equation}
Q_{1}=62.5\; {\rm GeV}, Q_{2}=29.1\; {\rm GeV}, Q_{3}=127.0\; {\rm GeV}
\end{equation}
for $\mu_{r}^{\rm init}=M_{H}=126$ GeV. It is noted that the expressions for $Q_{1}$ in Eq.(\ref{pmcq1}) and Eq.(\ref{seblmscale1}) are equal to each other at the LO and NLO level, we obtain almost the same value for $Q_{1}$ under the $R_\delta$-scheme and the seBLM approach.

We present a comparison of those three approaches in Table \ref{tableoth}, which shows that the pQCD convergence of the perturbative series clearly. Here as a comparison, the estimations for conventional scale setting and the results for the BKM scheme~\cite{BKM01} and the fractional analytic perturbation theory (FAPT) scheme~\cite{mfapt} are also presented.

At the present considered estimation up to four-loop levels all of those schemes including the conventional scale setting lead to good pQCD convergence. Especially, the PMC-I scheme, the $R_\delta$-scheme and the seBLM scheme have almost the same pQCD series. This is reasonable, since those three scale-setting schemes are designed to deal with the $\{\beta_i\}$-series of the process.

\begin{table}[ht]
\begin{tabular}{|c||c|c|c|}
\hline
& \multicolumn{3}{c|} {~~$\Gamma_{\rm NLO}$ (KeV)~~} \\
\hline
~~$\mu^{\rm init}_{r}$~~ & ~~$M_{H}/2$~~ & ~~$M_{H}$~~ & ~~$2 M_{H}$~~ \\
\hline
Conventional scale setting & 435.42 & 391.73 & 356.18 \\
\hline
seBLM~\cite{seblm} & 435.95 & 435.95 & 435.95\\
\hline
PMC-I~\cite{pmc2} & 435.03 & 435.99 & 436.06\\
\hline
$R_\delta$-scheme~\cite{pmc11} & 436.12 & 436.23 & 436.32\\
\hline
\end{tabular}
\caption{Initial scale dependence for $\Gamma_{\rm NLO}$ of $H\rightarrow b\bar{b}$. Here $\Gamma_i$ stands for the decay width at each perturbative order with $i={\rm LO}$, ${\rm NLO}$ and etc. The predictions of the PMC-I, $R_\delta$ and seBLM schemes are almost independent of $\mu^{\rm init}_{r}$. The cases for higher order decay widths $\Gamma_{\rm N^2LO}$, $\Gamma_{\rm N^3LO}$ and $\Gamma_{\rm N^4LO}$ are the similar. $M_{H}=126$ GeV. }
\label{scaleun}
\end{table}

In addition, one will observe that after eliminating the non-conformal $\{\beta_i\}$-series, one may also derive the (initial) renormalization scale independence for a fixed order prediction for those approaches. The initial scale dependence for the PMC-I, $R_\delta$-scheme and seBLM approaches on the NLO decay widths $\Gamma_{\rm NLO}$ are presented in Table \ref{scaleun}, where there typical initial scales $\mu^{\rm init}_{r}=M_H/2$, $M_H$ and $2 M_H$ are adopted. It shows clearly that the value of $\Gamma_{\rm NLO}$ are almost unchanged with $\mu^{\rm init}_{r}$. The higher-order terms have similar properties.

\section{summary} \label{sum}

The conventional scale setting procedure assigns an arbitrary range and an arbitrary systematic error to fixed-order pQCD predictions. And its error analysis can only get a rough estimation of the $\beta$-dependent nonconformal terms, not the entire perturbative series. As a possible solution, the PMC provides a systematic way to set the optimized renormalization scales for high energy processes. In principle, the PMC needs an initial value to initialize renormalization scale and renormalization procedures. It is found that the estimates of PMC are to high accuracy independent of the initial renormalization scale; even the PMC scales themselves are in effect independent of the initial renormalization scale and are `physical' at any fixed order. This is because the PMC scale itself is a perturbative series and those unknown higher-order $\{\beta_i\}$-terms will be absorbed into the higher-order terms of the PMC scale, which is strongly exponentially suppressed. Since the renormalization scale and scheme ambiguities are removed, the PMC can improve the precision of tests of the Standard Model and enhances the sensitivity to new phenomena. It is noted that the PMC applies the known non-conformal $\{\beta_i\}$-terms in a strict and systematic way to determine the behavior the coupling constant at each perturbative order. It provides an accurate estimation for the known perturbative series, and one may still need higher order calculations so as to known even higher-order conformal contributions, especially, when the perturbative series is not converge enough.

The PMC can be applied to a wide-variety of perturbatively-calculable collider and other processes. In addition to previous examples done in the literature, following its standard $R_\delta$-scheme procedures, we have done a through analysis of these two processes up to four-loop and three-loop levels. A comparison of the estimations under three approaches, i.e. the PMC-I approach, the $R_\delta$-scheme and the seBLM approach, have also been presented. We observe,
\begin{itemize}
\item Under conventional scale setting, it is often argued that by finishing a higher enough perturbative calculation, one can get desirable convergent and scale-invariant estimations. For the present considered channels, when considering up to three-loop level or four-loop level, as shown by FIGs.(\ref{Plot:hbb},\ref{Plot:hgg}), the decay width becomes almost invariant within the region of $\mu^{\rm init}_r \in [M_H/2,4M_H]$. However, even if a proper choice of $\mu^{\rm init}_r$ may lead to a value close to the experimental data by using conventional scale setting, this is a guess work and may not be the correct answer for a fixed-order estimation. Especially, if a process does not converge enough, one has to finish a more and more complex loop calculations so as to achieve the precision goal.

\item As shown by FIGs.(\ref{Plot:hbbpmc},\ref{Plot:hggpmc}), after PMC scale setting, the total decay widths of $H\rightarrow b\bar{b}$ and $H\to gg$ show a fast trend of stabilization versus the change of initial renormalization scale, which are almost flat even at the NLO level. The residual scale dependence due to unknown higher-order $\{\beta_i\}$-series has been greatly suppressed. This indicates that the PMC scale setting do provide a principle for setting the optimal renormalization scale of high energy processes.

\begin{figure}[t]
\includegraphics[width=0.50\textwidth]{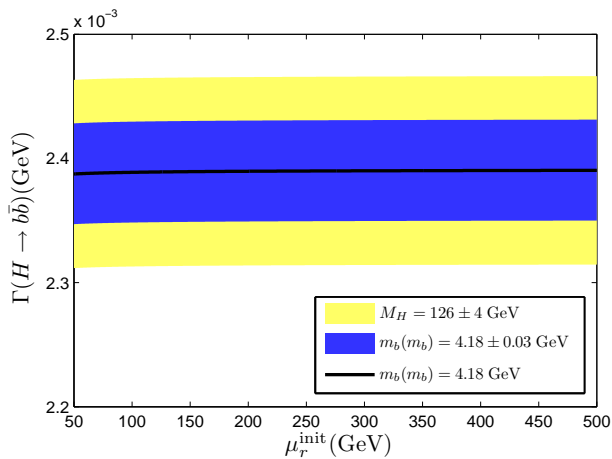}
\caption{Uncertainties for the decay width of $H\rightarrow b\bar{b}$ up to four-loop level. Two shaded bands shows the uncertainties from $M_H=126\pm 4$ GeV and the $\overline{\rm MS}$ $b$-quark mass $m_b(mb)=4.18\pm0.03$ GeV, respectively. The solid line is obtained by set the parameters to be their central values. } \label{Plot:pmcmu}
\end{figure}

\item In comparison to the previous estimations, e.g. Refs.\cite{newconv1,newconv2}, a more accurate predications for those two Higgs decay processes can be obtained. After PMC scale setting, we obtain the total decays widths for those two channels:
    \begin{eqnarray}
     \Gamma(H\rightarrow b\bar{b})&=& 2.389\pm0.073 \pm0.041\; {\rm MeV},\\
     \Gamma(H\rightarrow gg)&=& 0.373\pm0.030 \; {\rm MeV}.
    \end{eqnarray}
    where the first error is caused by varying $M_H$ with the region of $[122, 130]$ GeV, and the second error for the $H\to b\bar{b}$ channel is caused by varying the $\overline{\rm MS}$ running mass $m_b(mb)$ with the region of $[4.15, 4.21]$ GeV. For clarity, we put the uncertainty analysis for the $H\to b\bar{b}$ channel in FIG.(\ref{Plot:pmcmu}).

\item The $\beta$ function governs the running behavior of the coupling constant via the renormalization group equations, thus any approach that can properly identify the $\{\beta_i\}$-series for a physical observable will surely lead to equivalently the same predictions. Practically, one usually calculate the $n_f$-terms by considering the vacuum polarization contributions. After a certain scale setting, different ways of identifying $n_f$-series to $\{\beta_i\}$-series may lead to different effective $\{\beta_i\}$-series at each known perturbative order, different residual $\{\beta_i\}$-dependence because of unknown perturbative orders, or different pQCD convergence. A comparison of $H\to b\bar{b}$ for the PMC-I approach, the $R_\delta$-scheme and the seBLM approach has been presented in Table \ref{tableoth}. At the four-loop level all those approaches lead to good pQCD convergence, they have almost the same pQCD expansion series, and all of them are almost independent on the wide choice of the initial renormalization scale. This shows these three approaches are equivalent to each other. The residual differences of these approaches are caused by the unknown $\{\beta_i\}$-terms that could be suppressed to a required accuracy by finishing a more higher-order calculation.

\item As one subtle point, one may meet the problem of quite small (or near the fixed point) effective scales for a specific scale setting method. For example, we have noted that for the case of $R(e^+ e^-\to {\rm hadron})$ at the measured scale $Q$, we can obtained a convergent and precise conformal series up to four-loop level by applying the $R_\delta$-scheme, whose LO, NLO and NNLO PMC scales are~\cite{pmc11}: $Q_{1}=1.3Q$, $Q_{2}=1.2Q$ and $Q_{3}=5.3Q$, respectively. In contrast, by using the seBLM scheme, we shall obtain $\ln(Q^2/Q^2_2) \sim 167$~\cite{seblm}, which leads to quite small $Q_2$ out of pQCD domain. If, as suggested by PMC, we only deal with the $n_f$ series that rightly controls the running behavior of the coupling constant into the coupling constant via the standard way of seBLM, then we shall obtain more moderate seBLM scales, $Q_1=1.3Q$, $Q_2=1.1 Q$ and $Q_3=228.9Q$. And, similar to the present Higgs decays, we can obtain consistent results for $R(e^+ e^-)$ under both the seBLM and the PMC scale settings. Moreover, it is noted that such situation could be softened to a certain degree by using the commensurate scale relation~\cite{relation}, or one may solve it by using proper running behavior of the coupling constant in lower scale region~\cite{zhang}.

\item As another subtle point, even if one can eliminate the scale dependence at lower perturbative order as NLO, it may necessary to know the higher-order conformal contributions if we want to achieve even higher precision. Taking the case of $H\to b\bar{b}$ as an example, its N$^2$LO terms improves the NLO estimation by about $2\%$ and the N$^3$LO terms improves the N$^2$LO estimation by about $1\%$. The unknown higher-order non-conformal contributions can be roughly estimated by varying the final undetermined PMC scale as $Q_4$ via the conventional way, e.g. $[Q_4/2,2Q_4]$. If after PMC scale setting the final terms at a certain fixed order give negligible contribution, then we shall obtain quite accurate estimations at such fixed order.

    As shown by Table \ref{scaleun}, it is noted that by setting $\mu^{\rm init}_{r}=M_H/2$, the $H\to b\bar{b}$ NLO decay width $\Gamma_{\rm NLO}$ under the conventional scale setting is close to the PMC estimations. In this sense, a choice of $\mu_{r}\equiv M_H/2$ is better than the choice of $\mu_{r}\equiv M_H$ for the conventional scale setting. In fact, under such choice, one can also obtain a more convergent pQCD series for the conventional scale setting.

\item As a minor point, taking $H\to b\bar{b}$ as an example, we point out a wrong way of estimating the conventional scale error. The correct way is to set an initial scale $\mu^{\rm init}_r$ and get the full $\mu^{\rm init}_r$-dependent expression (\ref{hbbbt}), e.g. those terms proportional to $\ln(\mu^{\rm init}_r/M_H)$ are kept, and then by varying $\mu^{\rm init}_r \in[M_H/2,2M_H]$ to discuss its scale error. In this way, we have shown that the conventional scale error up to four-loop level is almost eliminated, as shown in FIG.(\ref{Plot:hbb}). The wrong way is to adopt the expression (\ref{hbbnf}) as the starting point, and directly varies the scale of coupling constant from $M_H$ to $M_H/2$ or $2M_H$ to discuss the scale error. In this wrong treatment, the log terms involving $\mu^{\rm init}_r \neq M_H$ disappear, which however may have sizable contributions. In fact, such a naive treatment shows the conventional scale error is still about $\pm 2\%$~\cite{newconv1,newconv2} for varying the scale within the region of $[M_H/2,2M_H]$ even at the four-loop level.

\end{itemize}

\hspace{1cm}

\noindent{\bf Acknowledgments}: We thank Stanley Brodsky, Matin Mojaza and Andrei L. Kataev for helpful discussions. This work was supported in part by Natural Science Foundation of China under Grant No.11075225 and No.11275280, by the Program for New Century Excellent Talents in University under Grant No.NCET-10-0882, and by the Fundamental Research Funds for the Central Universities under Grant No.CQDXWL-2012-Z002.

\end{document}